# How to Characterize Thermal Transport Capability of 2D Materials Fairly? – Sheet Thermal Conductance and the Choice of Thickness

Xufei Wu,[a] Vikas Varshney,[b,c] Jonghoon Lee,[b,c] Yunsong Pang,[a] Ajit K. Roy,[b] Tengfei Luo [a,d]

a). Aerospace and Mechanical Engineering, University of Notre Dame, Notre Dame, IN 46556

b). Materials and Manufacturing Directorate, Air Force Research Laboratory, Wright-Patterson Air Force Base, OH 45433

c). Universal Technology Corporation, Dayton, OH, 45342

d). Center for Sustainable Energy at Notre Dame, Notre Dame, IN 46556

**Abstract:**

Ever since the discovery of the record-high thermal conductivity of single layer graphene, thermal transport capability of monolayer 2D materials has been under constant spotlight. Since thermal conductivity is an intensive property for 3D materials and the thickness of 2D materials is not well defined, different definitions of thickness in literature have led to ambiguity towards predicting thermal conductivity values and thus in understanding the heat transfer capability of different monolayer 2D materials. We argue that if conventional definition of thermal conductivity should be used as the quantity to compare the heat transfer capability of various monolayer 2D materials, then the same thickness should be used. Alternatively, to circumvent the problem of ambiguous thickness completely, we also suggest that a "sheet thermal conductance" to be defined as an intensive 2D material property when characterizing the heat transfer capability of 2D materials. When converting literature thermal conductivity values of monolayer materials to this new property, some new features that were not displayed when using different thicknesses show up.

In bulk three dimensional (3D) materials, the amount of material that participates in heat transfer is directly proportional to the cross-sectional area (width × thickness) perpendicular to heat flow direction. The heat transfer capability in such cases is usually characterized in terms of thermal conductivity ($\kappa$), which is defined according to the Fourier's law as follows:[1]

$$\kappa = -\left(\dot{Q}/(w \times t)\right)/\nabla T \qquad (1)$$

where $\dot{Q}$ is the rate of heat transfer (energy per unit time), $w$ is the width and $t$ is the thickness in the cross-sectional plane, and $\nabla T$ is the temperature gradient along the heat flow direction. As it is defined, thermal conductivity is an intensive property, i.e., it does not depend on the mass or volume of the material participating in heat transfer.

As the material systems go from macroscopic length scales to extremely thin films, and to monolayer two dimensional (2D) materials, where the monolayer thickness $t$ is not unambiguously defined, it becomes an important question to ask if Eq. (1), which involves thickness as one of the important parameters, is the most appropriate approach to determine the heat transport characteristics of monolayer 2D systems. This ambiguity of thickness in monolayer 2D materials has become even more relevant in recent years because of theoretical predictions of several stable monolayer materials encompassing a wide range of elements.[2,3] Here, we believe that an alternate way of characterizing heat transport capability, similar to the sheet resistance in electronics, can proposed to remove the ambiguity of monolayer thickness in predicting heat transport capability of 2D materials.

Ever since Balandin et al.[4] reported the record-high thermal conductivity of single layer graphene (4840-5300 W/mK), the thermal transport properties of 2D materials have been under constant research spotlight.[5-12] Based on the extremely high thermal conductivity measured in experiments, the heat transfer capability of graphene is often compared to that of diamond – the most thermally conductive known material in 3D form with the thermal conductivity of 2200 W/mK at room temperature.[13,14] Graphene from some experiments have reported higher thermal conductivity values (> 2200 W/mK), suggesting it to be the most thermally conductive material.[15] This, however, is a tricky comparison because to describe the heat transfer capability of a 2D material using conventional definition of thermal conductivity (Eq. (1)), a value for thickness needs to be chosen. For their estimation, Balandin et al.[4] used the interlayer distance between each layer in graphite as the "thickness" of graphene when calculating the thermal conductivity. This was an agreeable and also necessary choice when trying to make sense of the heat transfer capability of a 2D material using thermal conductivity, especially in comparison to its 3D counterpart – graphite. The rationale can be understood like this: Bulk graphite can be regarded as a stack of graphene. Since thermal conductivity is a intensive property, which does not depend on the actual thickness of the material, it is intuitive to assign the total thickness of graphite divided by the number of layers in the stack as the "nominal" thickness of a single layer graphene, which come out to be the interlayer distance of 3.35 Å. The situation is the same for multi-layer graphene, where the thickness is usually assigned as 3.35 Å multiplied by the number of layer when calculating their thermal conductivity.[16-18] Actually, the thermal conductivity values from such calculations are manifestations of the heat transfer capability of a layer of the 2D material when it is in the multi-layer or bulk.

From a notably different perspective, however, it can be also argued that since graphene is never perfectly flat in the monolayer form because of its out-of-plane vibrational modes which are often populated at operational temperatures, whether the interlayer distance as a measure of thickness is completely unambiguous or should there be any modification to such a value? Such a problem was discussed for carbon nanotube, whose effective diameter was shown to increase with temperature due to thermal vibration.[19] Such consideration is meaningful and necessary when there are many layers of 2D materials stacked together or many carbon nanotubes bundled together.

However, the question of "what thickness should one use" becomes trickier when comparing the heat transfer capability among different 2D materials using thermal conductivity. In most such comparisons, the interlayer distances of the corresponding 3D materials, as illustrated as $t_1$ in Fig. 1a, are often used.[9,12,20] However, this is not the only definition that has been used in the reported literature. Researchers have also used the buckling distance – the out of plane direction distance between the topmost and bottommost atoms in the 2D structure – as the thickness ($t_2$ in Fig. 1).[21] Others have used the summation of the buckling distance and the van der Waals (vdW) radii of the outer-most surface atoms ($t_3$ in Fig. 1).[10,22] While the latter choice ($t_3$) usually yields values close to the inter-layer distance ($t_1$), the second choice has an obvious flaw – the graphene thickness would be zero since it is of single atom layer. Because the calculated thermal conductivity is inversely proportional to the chosen thickness, the quantitative value of thermal conductivity varies strongly with the thickness chosen. It should be understood by now that no matter which definition is used, the thickness is merely a numerical factor so that a thermal conductivity value can be calculated using Eq. (1). Hence, we argue (as detailed below) that if the "thermal conductivity" definition is used to compare the heat transport capability of 2D materials, the same thickness value should be used, and better yet, we suggest an alternate observable to characterize heat transport capability of 2D materials, which does not require the concept of thickness. Ideally, for the comparison of heat transfer capability among 2D materials, an intensive 2D property independent of the chosen thickness should be adopted.

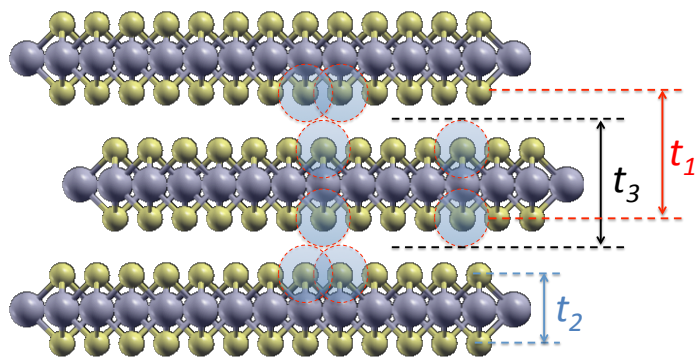

**Figure 1**. *Different definitions of thicknesses. $t_1$ – interlayer distance; $t_2$ – buckling distance; $t_3$ – buckling distance plus the sum of the vdW radii of outer-most atoms.*

Here, we first justify the reason of using the same thickness value when comparing the thermal conductivity of different 2D materials. This can be best illustrated through the following thought experiment: For a 2D material involved in thermal transport between a heat source and a heat sink, no matter how "thick" it is, all the thermal energy has to pass through it. Thus, the amount of material that participates in heat transfer is only proportional to the width of the 2D material, but has nothing to do with the "thickness". To further argue this point, we schematically show two 2D materials with different "thicknesses" in Fig. 2. We assume these two materials have the same width and let them bridge the same heat source ($T_H$) and heat sink ($T_L$). If the rates of heat transfer ($\dot{Q}$) are the same for these two cases, then it is natural to say that these two materials have the same "heat transfer capability".

If we were to use a quantity to characterize the heat transfer capability of a 2D material, in analogous to the 3D case (Eq. (1)), it would be defined as:

$$\kappa' = -\left(\dot{Q}/w\right)/\nabla T \quad (2)$$

noting that thickness should not be part of the definition in characterizing the heat transfer capability of 2D materials. For the cases presented in Fig. 2, we will obtain the same "thermal conductivity" if we use the definition given by Eq. (2), but not according to the definition for the 3D case (Eq. (1)) due the different "thickness" of the materials. In summary, Eq. (2) defines the 2D analog of the intensive property (thermal conductivity) defined in Eq. (1) to characterize the intrinsic heat transfer capability of a 2D material.

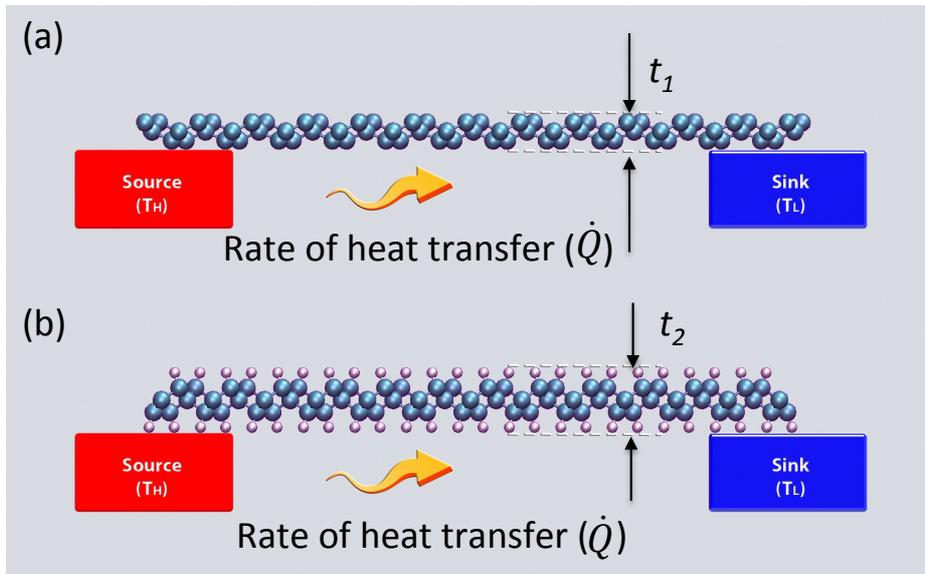

**Figure 2**. *Illustration of the heat transfer capability of 2D materials which is independent of*

*thickness.*

However, since the "thermal conductivity" defined by Eq. (1) is so widely adopted in the scientific community, it is frequently used to describe the heat transfer capability of 2D materials by choosing certain value of 2D materials' "thickness", which, as pointed out previously, is merely a numerical factor (or a unit converter) to retrieve the conventional 3D definition of the thermal conductivity. We argue that either such a numerical factor should be the same for all materials being compared, i.e., the same thickness should be used to avoid biased evaluation of heat transfer capability of one 2D material over another, or an alternative definition (such as shown in Eq. (2)) should be used to characterize two-dimensional heat transfer. Unless there is a unified value for thickness, it is inevitable that different thicknesses will be used for different 2D materials in different studies. Moreover, it is also inconvenient to have to convert reported thermal conductivity values from one study to another due to the differences in thickness adopted in different literatures.

In Fig. 3, we have taken from the literature a number of thermal conductivity values from first-principles lattice dynamics calculations for various 2D materials and converted them using the thickness of the interlayer distance of graphite (3.35 Å) for comparison for their heat transfer capability. This list is not meant to be inclusive of all existing literature values as there are a large number of very informative studies using other methods such as molecular dynamics simulations or lattice dynamics with empirical potentials as well.[22-26] It is apparent in Fig. 3 that converting the thermal conductivity values has led to some key observations that was not evident in the literature values. For example, the heat transfer capability of 2D $WS_2$ is actually very similar to that of the fluorographene as indicated by the converted thermal conductivity values. In additions, all the 2D transition metal dichalcogenides and IV-VI materials appear to be more thermally conductive than previously indicated.

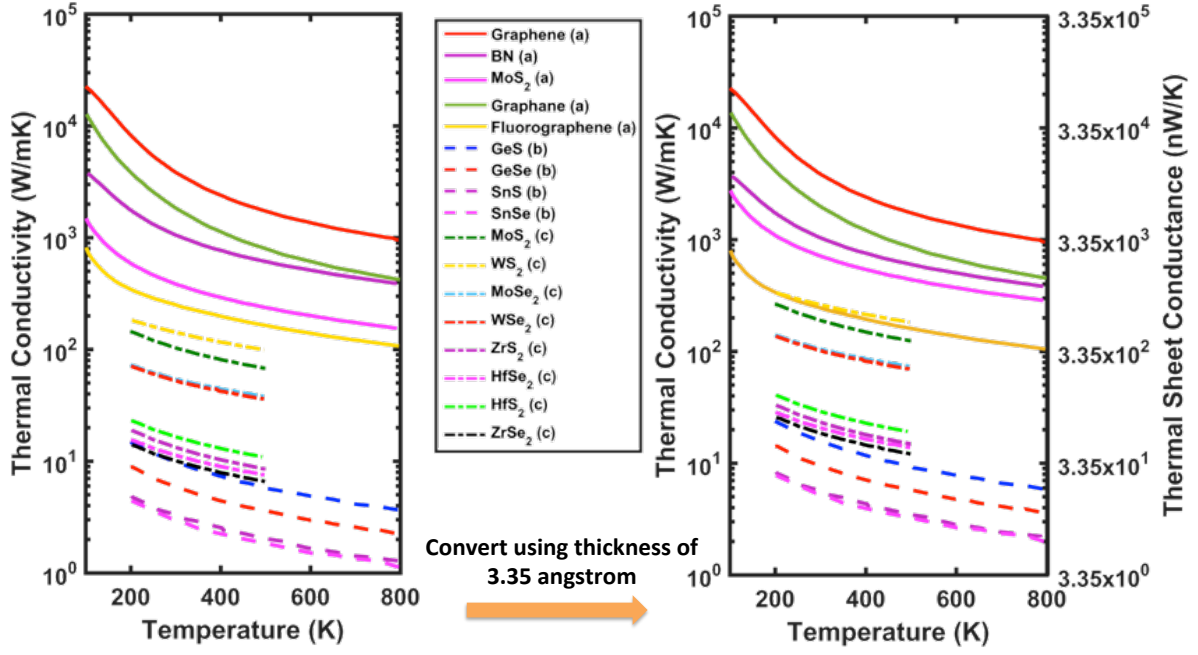

*Figure 3*. *Thermal conductivity values from literature (left) and those converted using the same thickness (3.35 Å) and thermal sheet resistance (right). (a)- Ref. [9]; (b)- Ref. [12]; (c)- Ref. [20].*

In fact, towards the understanding of the thermal transport capability of 2D materials, Eq. (2) can entirely avoid the dilemma of "thickness" all together as it does not involve thickness in this definition. We name the quantity *(κ')* as "thermal sheet conductance". It is called "conductance" because it has the unit of thermal conductance (W/K), but unlike the 3D thermal conductance, which depends on the actual dimension of the specimen, thermal sheet conductance is an intensive property for 2D materials. The term "thermal sheet conductance" is inspired by a similar property defined for 2D electron transport called "sheet resistance" as discussed below.[27]

Analogous to thermal conductivity from Fourier's law as defined in Eq. (1), electrical conductivity $\sigma$, according to the generalized Ohm's law,[28] can be defined as:

$$\sigma = J/E = \left(I/(w \times t)\right)/E \qquad (3)$$

Here, *J* is current density, which is total current, *I*, divided by the cross sectional area (width (*w*) × thickness (*t*)), and *E* is the electric field. The sheet resistance, $R_s$, is formally defined as:

$$R_s = \frac{1}{\sigma \cdot t} \qquad (4)$$

Combining Eqs. (3) and (4), we can obtain the following expression:

$$R_s^{-1} = (I/w)/E \qquad (5)$$

where $R_s^{-1}$ can be defined as the "sheet conductance". Eq. (5) shows that the sheet conductance does not depend on the thickness of the materials. It can be observed that Eq. (2) is perfectly analogous to Eq. (5). We rewrite Eq. (2) using a new symbol $G_s$ to denote thermal sheet conductance:

$$G_s = -(\dot{Q}/w)/\nabla T \qquad (6)$$

Thermal sheet conductance can be converted into thermal conductivity if a thickness, $t$, is to be further defined:

$$\kappa = G_s / t \qquad (7)$$

In Fig. 3, we have also converted all thermal conductivity values into thermal sheet conductance by dividing the thermal conductivity values by the reported thickness. We argue that when using the newly defined thermal sheet conductance to characterize the heat transfer capability of 2D materials, their heat transfer capability can be compared more fairly without the need of selecting an appropriate thickness. For convenience, we have also compiled the room temperature data of different 2D materials into Table 1.

The newly defined thermal sheet conductance will provide a fair comparison between different monolayer 2D materials. In addition, this property can also be utilized for comparing the heat transfer capabilities of monolayer 2D materials and their multi-layer counterparts. Such comparisons have been seen for a number of 2D materials,[18,29-31] especially graphene. If the comparison is between the "true" heat transfer capability of monolayer and multi-layer 2D materials, then Eq. (6) should be used directly. One would imagine that as the number of layers increases, $G_s$ will increase accordingly. This is actually the right description of the heat transfer capability because when more layers participate in heat transfer, the material can indeed allow larger rate of heat transfer at a given temperature gradient. However, if one wants to compare the heat transfer capability of a 2D material in the monolayer form and that of a single layer when it is in a multi-layer stack, which is actually the comparison made in the literature when thermal conductivity values are used,[18,29-31] one can simply divide the thermal sheet thermal conductance by the number of layers.

In summary, we have discussed the ambiguities and difficulties in using thermal conductivity defined by Fourier's law to characterize the heat transfer capability of 2D materials. We have shown that when comparing the heat transfer capability of different 2D materials using thermal conductivity, the same thickness value should be used. We have also suggested a newly defined observable, thermal sheet conductance, for characterizing the heat transfer capability of 2D materials. This quantity is an intensive property for 2D materials and does not involve thickness.

**Table 1.** Thermal conductivity and thermal sheet conductance of different 2D materials and percentile reduction compared to graphene. (a)- Ref. [9]; (b)- Ref. [12]; (c)- Ref. [20].

| Material | Thermal Conductivity (W/mK) | Percentile reduction compared to graphene | Thermal Conductivity using the same thickness (W/mK) | Thermal sheet conductance (nW/K) | Percentile reduction compared to graphene |
|---|---|---|---|---|---|
| Graphene | 3846[a] | 0% | 3846[a] | 12884[a] | 0% |
| BN | 1055[a] | -72.6% | 1037[a] | 3474[a] | -73.0% |
| Graphane | 1834[a] | -52.3% | 1970[a] | 6600[a] | -48.8% |
| Fluorographene | 250.4[a] | -93.5% | 245.1[a] | 821.1[a] | -93.6% |
| GeS | 9.873[b] | -99.7% | 15.80[b] | 52.93[b] | -99.6% |
| GeSe | 5.890[b] | -99.9% | 9.426[b] | 31.58[b] | -99.8% |
| SnS | 3.269[b] | -99.9% | 5.576[b] | 18.68[b] | -99.9% |
| SnSe | 2.980[b] | -99.9% | 5.238[b] | 17.55[b] | -99.9% |
| $MoS_2$ | 385.5[a], 103.4[c] | -90.0%, 97.3% | 713.9[a], 189.8[c] | 2391[a], 635.8[c] | -81.4%, 95.1% |
| $WS_2$ | 141.9[c] | -96.3% | 261.0[c] | 874.4[c] | -93.2% |
| $MoSe_2$ | 54.21[c] | -98.6% | 104.7[c] | 350.8[c] | -97.3% |
| $WSe_2$ | 52.47[c] | -98.6% | 101.5[c] | 340.0[c] | -97.4% |
| $ZrS_2$ | 13.31[c] | -99.6% | 23.25[c] | 77.89[c] | -99.4% |
| $HfSe_2$ | 11.30[c] | -99.7% | 20.71[c] | 69.38[c] | -99.5% |
| $HfS_2$ | 16.56[c] | -99.6% | 29.06[c] | 97.35[c] | -99.2% |
| $ZrSe_2$ | 10.10[c] | -99.7% | 18.55[c] | 62.14[c] | -99.5% |

**Acknowledgements:**

X.W. and T.L. thanks the support by NSF (1433490). T.L. thanks the support from the Air Force Summer Faculty Fellowship.